\documentclass[aps,prl,twocolumn,groupedaddress]{revtex4}
\usepackage{graphicx}
\usepackage{helvet}
\usepackage{amsmath,amssymb}
\usepackage{bm}

\begin{document}

\title{Separation of Charge Instability and Lattice Symmetry Breaking in an Organic Ferroelectric}

\author{Ryosuke Takehara$^{1}$ }
\author{Keishi Sunami$^{1}$ }
\author{Fumitatsu Iwase$^{1}$ }
\thanks{Current address: Department of Physics, Tokyo Medical University, Shinjuku-ku, Tokyo 160-8402, Japan}
\author{Masayuki Hosoda$^{1}$ }
\author{Kazuya Miyagawa$^{1}$ }
\author{Tatsuya Miyamoto$^{2}$ }
\author{Hiroshi Okamoto$^{2,3}$}
\author{Kazushi Kanoda$^{1}$ }
\thanks{Corresponding author, e-mail:kanoda@ap.t.u-tokyo.ac.jp}

\affiliation{
$^{1}$ Department of Applied Physics, University of Tokyo, Bunkyo-ku, Tokyo, 113-8656, Japan \\
$^{2}$ Department of Advanced Materials Science, University of Tokyo, Kashiwa, Chiba, 277-8561, Japan \\
$^{3}$ AIST-UTokyo Advanced Operando-Measurement Technology 
Open Innovation Laboratory (OPERANDO-OIL), 
National Institute of Advanced Industrial Science and Technology (AIST), Chiba 277-8568, Japan}

\date{\today}

\begin{abstract}
	We investigate the charge and lattice states in a quasi-one-dimensional organic ferroelectric 
material, TTF-QCl$_{4}$, under pressures of up to 35 kbar 
by nuclear quadrupole resonance experiments. 
	The results reveal a global pressure-temperature phase diagram, 
which spans the electronic and ionic regimes of ferroelectric transitions, 
which have so far been studied separately, in a single material. 
	The revealed phase diagram clearly shows that the charge-transfer instability and 
the lattice symmetry breaking, which coincide in the electronic ferroelectric regime 
at low pressures, bifurcate at a certain pressure, leading to the conventional ferroelectric regime. 
	The present results reveal that the crossover from electronic to ionic ferroelectricity 
occurs through the separation of charge and lattice instabilities.
\end{abstract}



\maketitle

\begin{figure}
	\includegraphics[width=8.6cm]{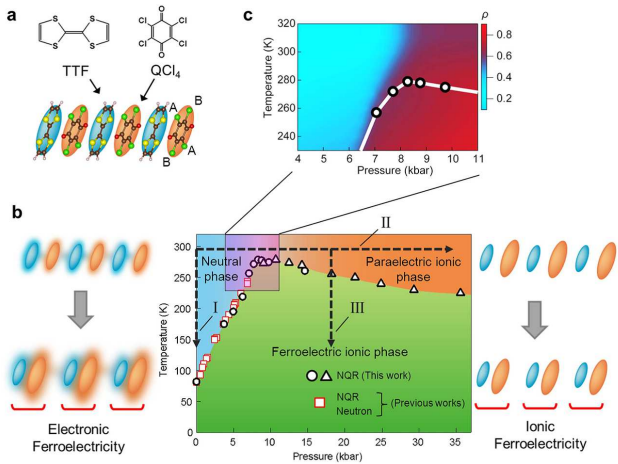}
	\caption{\label{Fig1} (Color online) 
Structure and pressure-temperature phase diagram of TTF-QCl$_{4}$. 
(a) One-dimensional alternating stack of TTF and QCl$_{4}$ molecules in the neutral state.
(b) Pressure-temperature phase diagram of TTF-QCl$_{4}$ determined by NQR experiments 
(black open circles and black triangles); the former is obtained from sample $\#$1 studied 
under fine pressure control, whereas the latter is from sample $\#$2 studied in 
a wide pressure range up to 35 kbar. 
Red open squares indicate previous results obtained from NQR and neutron experiments 
\cite{Cailleau_1997}. 
The ferroelectricity is dominated by the modification of the electronic wave at 
low pressures (left figure), whereas it is caused by an ionic displacement at high pressures 
(right figure). 
The size of the ellipsoid schematically expresses the total electron density in the HOMO (highest occupied molecular orbital) and LUMO (lowest unoccupied molecular orbital). 
(c) Profile of the degree of charge transfer, $\rho$, defined as TTF$^{+ \rho}$ QCl$_{4}^{-\rho}$, 
in the squared region of the $P$-$T$ phase diagram in (b). 
The data of sample $\#$1 studied under fine pressure tuning are used.}
\end{figure}

	There has been increasing interest in electronic ferroelectricity, 
which stems from deformation of the electronic wave function coupled 
with lattice symmetry breaking, invoking a Berry-phase description of 
electronic polarization beyond the conventional notion of ionic displacement \cite{Resta_2007}. 
	How does a ferroelectric transition of electronic nature cross over to a conventional one 
of ionic nature ? 
	This is a fundamental issue of ferroelectricity; however, it is yet to be answered because 
these two regimes of ferroelectricity have so far been studied with separate materials. 
	The organic charge-transfer crystal TTF-QCl$_{4}$ (tetrathiafulvalene-p-chloranil) 
\cite{Cointe_1995, Kobayashi_2012} is a candidate system for solving this missing link 
in ferroelectricity research. 
	TTF-QCl$_{4}$ is composed of one-dimensional alternating face-to-face stacks of 
donor molecules, TTF, and acceptor molecules, QCl$_{4}$ (Fig.~\ref{Fig1}(a)) \cite{Mayerle_1979}, 
carries an emergent "electronic ferroelectricity" arising from an electronic displacement, 
which is distinguished from the conventional ionic displacement 
\cite{Kobayashi_2012, Giovannetti_2009, Ishibashi_2014}. 
	A key to understanding the novel ferroelectricity in TTF-QCl$_{4}$ is 
a neutral-to-ionic transition with charge transfer, 
which occurs when an increase of electronic occupation energy is surmounted by a gain of 
the Madelung energy under temperature or pressure variation \cite{Torrance_1981}. 
	At ambient pressure, TTF-QCl$_{4}$ is in a neutral state at room temperature 
but undergoes a charge-transfer transition to an ionic state at 81 K \cite{Torrance_1981}, 
simultaneously accompanied by a lattice dimerization forming the donor-acceptor (DA) pairs 
\cite{Cointe_1995, Girlando_1983, Tokura_1985}, 
which creates strongly charge-lattice-coupled ferroelectricity. 
	The resultant polarization, which the classical framework fails to explain, is well described in terms of the electronic Berry phase 
\cite{Giovannetti_2009, Ishibashi_2014, King-Smith_1993, Vanderbilt_1993, Resta_1994}.

%\newpage

%\newpage
\begin{figure}
	\includegraphics[width=8.6cm]{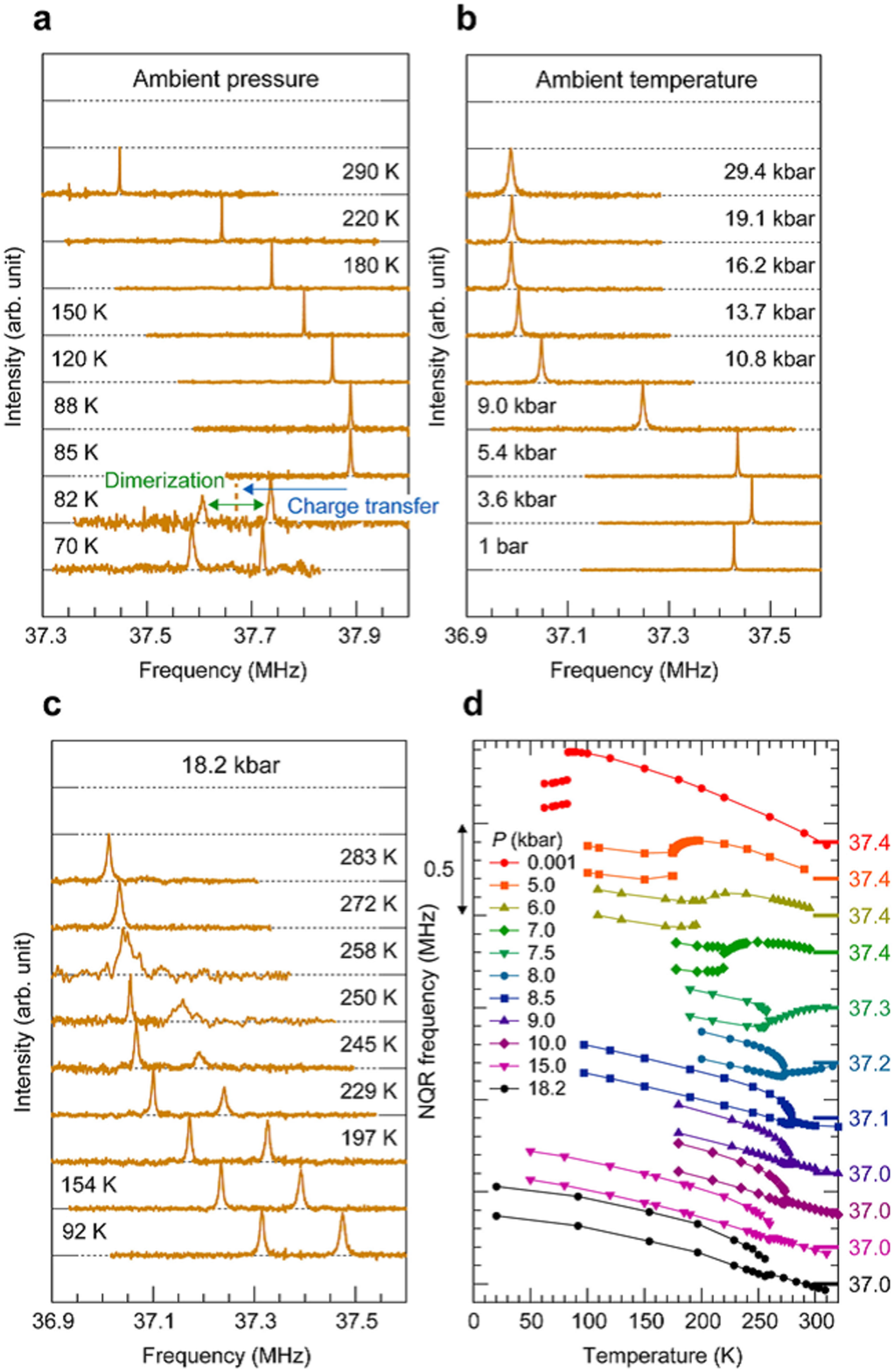}
%  \begin{center}
%	\includegraphics[width=16cm]{figure2.eps}
	\caption{\label{Fig2} (Color online)  NQR spectra and pressure-temperature profile of 
NQR frequency. 
a-c, $^{35}$Cl NQR spectra measured under temperature variation at ambient pressure 
(a), under pressure variation at ambient temperature 
(b), and under temperature variation at 18.2 kbar (c). 
The broken line in the 82 K spectrum in a represents the position of the centre-of-gravity 
of the spectrum. 
The line splitting to a doublet in a and c indicates the dimerization transition, 
which causes a lattice symmetry breaking. 
The abrupt shift in the centre-of-gravity of the spectrum signifies a charge transfer. 
d, NQR frequencies plotted against temperature for fixed pressures. 
The vertical scale (NQR frequency) is offset for each value of pressure for clarity; 
the reference values of frequency for each pressure are indicated on the right axis with 
the tick marks. 
The pressure values labelled in the panel are the room-temperature values.}
%  \end{center}
\end{figure}

	Earlier experiments performed under pressures of up to 9 kbar 
\cite{Takaoka_1987, Cailleau_1997, Cointe_2017} showed that 
the neutral-to-ionic transition occurs simultaneously with the dimerization transition 
(Fig.~\ref{Fig1}b). 
	However, there was an indication that the charge-lattice coupling becomes weak with 
increasing pressure \cite{Cointe_2017}, implying that the polarization is less electronic 
in origin but more ionic, namely, more classical, under pressure. 
	Thus, TTF-QCl$_{4}$ potentially serves as a platform for studying the link between 
the modern wave-like and classical particle-like pictures for ferroelectricity using pressure 
as a parameter. 
	To clarify this missing link, we have extended the pressure study of this material up to 
35 kbar by means of NQR, which probes the charge/lattice state.

	Charge transfer and lattice dimerization are signified by the profile of NQR lines, 
which appear at frequencies proportional to the electric-field gradient at the nuclear position, 
determined by the local charge distribution. 
	We performed $^{35}$Cl NQR measurements on two polycrystalline samples of TTF-QCl$_{4}$, 
which contains four Cl atoms in QCl$_{4}$ (Fig.~\ref{Fig1}a), under zero magnetic field. 
	The results for sample $\#$1 and $\#$2 are represented as open circles and triangles 
in Fig.~\ref{Fig1}b and \ref{Fig1}c, respectively. 
	NQR signals were acquired with the standard spin-echo techniques. 
	Hydrostatic pressure, $P$, was applied to the sample using 
a BeCu/NiCrAl dual-structured clamp-type cell with Daphne 7373 (P $<$ 20 kbar) and 
7474 (20 $<$ P $<$ 35 kbar) oils as the pressure media. 
	The solidification pressures of the Daphne 7373 and 7474 oils are 
approximately 22 and 37 kbar at room temperature, respectively, and thus, 
the hydrostaticity is maintained over the entire pressure range in this study. 
	The pressure, which was maintained by clamping the pressure cell at room temperature, 
was somewhat reduced by solidification of the oils when the sample is cooled. 
	This effect is corrected with reference to the reported calibration data \cite{Murata_1997}, 
in the pressure and temperature range where the effect is appreciable.

	Figure~\ref{Fig2}a shows the temperature evolution of a $^{35}$Cl NQR line measured 
at ambient pressure (along arrow I in Fig.~\ref{Fig1}b). 
	A single line in the neutral state at high temperatures splits into two lines 
when TTF-QCl$_{4}$ undergoes a ferroelectric transition at $\sim$82 K 
because the two A (B) Cl atoms in QCl$_{4}$ become inequivalent due to inversion symmetry 
breaking caused by DA dimerization \cite{Cointe_1995}. 
	The line splitting is accompanied by a clear shift of the centre-of-gravity of the lines, 
which signifies a change in molecular charge due to the neutral-to-ionic transition. 
	The simultaneous line splitting and shift demonstrate that the charge transfer transition 
and the dimerization transition coincide, as earlier results indicated 
\cite{Gourdji_1991, Gallier_1993}.

	Next, we traced the NQR line under a pressure sweep of up to 35 kbar at room temperature 
(along arrow II in Fig.~\ref{Fig1}b). 
	The NQR line shows a pronounced shift at $\sim$ 9 kbar (Fig.~\ref{Fig2}b), 
which locates on the extrapolation of the ferroelectric transition line indicated by 
earlier reports \cite{Cailleau_1997}. 
	This sharp line shift is attributable to the charge transfer 
as demonstrated by previous IR measurements \cite{Matsuzaki_2005}. 
	Remarkably, however, no line splitting was observed; 
namely, the sharp charge transfer that occurs at $\sim$ 9 kbar is not accompanied 
by the long-range order of lattice dimerization. 
	Next, we fixed the pressure at 18.2 kbar and monitored NQR spectra on cooling from 
room temperature (along arrow III in Fig.~\ref{Fig1}b). 
	A single line observed at room temperature splits below 258 K (Fig.~\ref{Fig2}c), 
accompanied by only a small spectral shift, indicating a dimerization transition with 
a vanishingly small charge transfer. 
	These results suggest that the charge transfer and dimer order are decoupled at high temperatures and pressures.

	To see the profiles of the charge transfer and the lattice instability in the $P$-$T$ plane, 
we examined the temperature dependence of NQR spectra at pressures that were increased 
in small steps. 
	The splitting of NQR frequency clearly identifies the dimerization transition 
for each pressure (Fig.~\ref{Fig2}d). 
	The plot of the transition points in Fig.~\ref{Fig1}b reveals a novel $P$-$T$ phase diagram 
in which the dimerization transition temperature unexpectedly ceases to increase with 
pressure at approximately 10 kbar and decreases at higher pressures.

\begin{figure}
	\includegraphics[width=8.6cm]{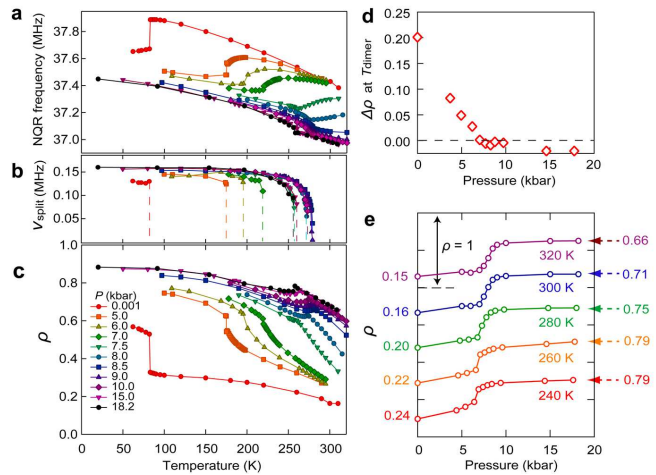}
	\caption{\label{Fig3} (Color online) Degree of charge transfer and lattice dimerization. 
\textbf{a}-\textbf{c}, Temperature dependences of NQR frequency $\nu_{\textrm{Q}}$ 
(a), NQR line splitting $\nu_{\textrm{split}}$ 
(b), the degree of charge transfer, $\rho$, 
(c), under various pressures. 
The magnitude of $\nu_{\textrm{split}}$ is a measure of the degree of dimerization. 
The value of $\rho$ is deduced from the value of $\nu_{\textrm{Q}}$ (Supplementary material). 
The pressure values labelled in the panel are the room-temperature values. 
d, The degree of charge transfer, $\rho$, plotted against pressure at fixed temperatures 
indicated in the panel. 
The data for 320 K are extrapolated values at each pressure. 
e, Discontinuity in $\rho$ at the dimerization transition, $\Delta\rho$, for each pressure. 
The pressure values of the horizontal axes in d and e are ones corrected for the temperature dependence of pressure (Supplementary material).}
\end{figure}

	The molecular charge and inversion symmetry breaking in QCl$_{4}$ are probed by 
the centre-of-gravity frequency of the NQR line, $\nu_{\textrm{Q}}$, 
and the line splitting, $\nu_{\textrm{split}}$, respectively, 
whose temperature dependences are shown in Figs.~\ref{Fig3}a and \ref{Fig3}b. 
	It is also known that $\nu_{\textrm{Q}}$ varies with temperature in a generic manner 
due to the thermally activated molecular motion, which effectively reduces the anisotropy of 
the quadrupole tensor due to motional averaging \cite{Bayer_1951, Koukoulas_1990}. 
	This effect on $\nu_{\textrm{Q}}$ can be separated from the effect of charge in 
question by treating the effect semi-empirically so that we can extract the degree of 
charge transfer, $\rho$, defined as D$^{+\rho}$A$^{-\rho}$, (Fig.~\ref{Fig3}c) from 
the $\nu_{\textrm{Q}}$ values (Fig.~\ref{Fig3}a) using the optical data 
\cite{Matsuzaki_2005, Horiuchi_2000} (Supplementary material).
At ambient pressure, $\rho$ is $\sim$ 0.2 at room temperature and, on cooling, 
gradually increases to $\sim$ 0.3 at 82 K, where it jumps above 0.5, coincidentally followed by the dimerization transition. 
	Under applied pressures, the jump in $\rho$, $\Delta\rho$, is diminished and 
nearly vanishes under $\sim$ 7.5 kbar (Fig.~\ref{Fig3}d). 
	At higher pressures, the $\rho$ values are well above 0.5, 
indicating that the system is in the ionic state even before the dimerization transition, 
and these features of $\rho$ are consistent with previous optical studies performed 
below 11 kbar \cite{Dengl_2014}. 
	At high pressures of 15 and 18.2 kbar, $\Delta\rho$ takes a slightly negative value 
(Fig.~\ref{Fig3}d), very probably because the largely transferred charge is backflowed by 
the molecular orbital hybridization between the DA pair promoted by the dimerization.

	Figure~\ref{Fig3}e displays the variation of $\rho$ with pressure at fixed temperatures. 
	At 240 and 260 K, $\rho$ shows sharp changes at pressures between 7.0 and 7.5 kbar, 
coinciding with the dimerization transition. 
	Remarkably, even at 280, 300 and 320 K, when no dimer order was observed, 
$\rho$ exhibits continuous but sharp changes in a narrow pressure range of 7.5 to 8.0 kbar. 
	The contour plot of the $\rho$ values in the $P$-$T$ phase diagram (Fig.~\ref{Fig1}c) 
illustrates that the charge transfer and the dimer order, which coincide at low pressures, 
are separated above $\sim$ 7.5 kbar. At $\sim$ 8.5-9 kbar, just above the bifurcation pressure, 
the line splitting, $\nu_{\textrm{split}}$, on the transition to the dimer order nearly 
vanishes (Fig.~\ref{Fig3}b), indicating that the ferroelectric transition becomes 
second order around this pressure region.

%\newpage

   The separation of the charge transfer and the lattice symmetry breaking means that 
an ionic phase without dimer order extends between the bifurcating lines 
(the red region in Fig.~\ref{Fig1}b). 
	This result appears to be incompatible with an earlier study using infrared spectroscopy, 
which detected the molecular vibrational modes characteristic of DA dimerization at pressures 
above 8 kbar \cite{Okamoto_1989}, where the NQR experiments find no symmetry breaking. 
	A key to a consistent understanding of the two seemingly inconsistent results is 
the time scale of the experimental probes. 
	NQR detects the electronic and lattice states on a time scale of $\sim$ 10$^{-7}$ s, 
whereas infrared spectroscopy captures their snapshot in the time domain of $\sim$ 10$^{-12}$ s. 
	If the DA dimers are not long-range ordered but fluctuating in between the two time scales, 
NQR spectra would yield a single line due to the motional narrowing 
whereas the infrared probe would detect the instantaneous dimerized states, 
which explains the two contrasting results coherently.

\begin{figure}
	\includegraphics[width=7.5cm]{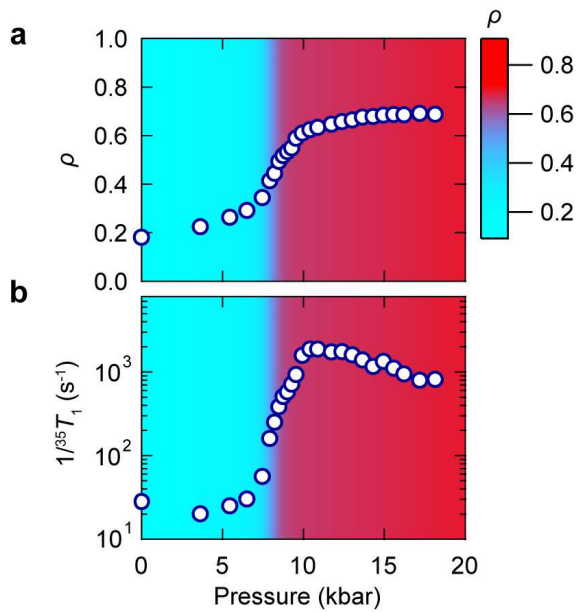}
	\caption{\label{Fig4} (Color online) Variation of the degree of charge transfer, 
NQR relaxation rate on crossover from a neutral state to an ionic dimer-liquid state. 
Pressure dependence of the degree of charge transfer $\rho$ 
(a), $^{35}$Cl NQR spin-lattice relaxation rate 1/$^{35}T_{1}$ 
(b) at room temperature. 
The $\rho$ values in \textbf{b} are estimated from the NQR frequency 
$\nu_{\textrm{Q}}$ (see Supplementary material for detail). 
The range of colour in a and b corresponds to the value of $\rho$ as in Fig. 1c.}
\end{figure}

	A further indication of the dimer liquid state is given by the dynamical properties of 
NQR spectroscopy performed under fine pressure variation at room temperature 
(along line II in Fig.~\ref{Fig1}b). 
	Figures~\ref{Fig4}a and \ref{Fig4}b show the pressure evolutions of 
the degree of charge transfer and $^{35}$Cl NQR relaxation rate 1/$^{35}T_{1}$, respectively, 
which probes the fluctuations of the electric-field gradient arising from fluctuations 
of charge and/or lattice. 
	At $\sim$ 8 kbar, a charge transfer from TTF to QCl$_{4}$ occurs sharply (Fig.~\ref{Fig4}a), 
and concomitantly, 1/$^{35}T_{1}$ increases (Fig.~\ref{Fig4}b), 
indicating that charge and/or lattice fluctuations are strongly enhanced in 
the transient pressure region of charge transfer. 
	At higher pressures, the charge transfer is nearly complete (Fig.~\ref{Fig4}a); 
however, the 1/$^{35}T_{1}$ values remain two orders of magnitude higher than in the neutral phase, 
indicating extraordinarily enhanced lattice fluctuations in the charge-transferred ionic phase, 
consistent with the dimer liquid picture. 
	Thus, it turns out that the new phase diagram of TTF-QCl$_{4}$ contains three distinct 
phases - a neutral phase, a paraelectric ionic phase and 
a ferroelectric ionic phase (Fig.~\ref{Fig1}b) - which correspond to the gas, 
liquid and solid phases of dimers with respect to the lattice degrees of freedom.

%\newpage

	The ferroelectric order of TTF-QCl$_{4}$ at ambient and low pressures is of 
electronic origin \cite{Kobayashi_2012, Giovannetti_2009, Ishibashi_2014}, which explains 
the large polarization observed in TTF-QCl$_{4}$, whereas the classical model of 
ionic displacement fails to explain it \cite{Kobayashi_2012}. 
	At high pressures well above 8 kbar, the charge transfer is nearly complete, meaning that 
the ferroelectricity is classical. 
	The phase diagram shown here (Fig.~\ref{Fig1}b) illustrates how the electronic 
ferroelectricity crosses over to the ionic one. 
	When multiple degrees of freedom are involved in a phase transition, 
it is generally first order. 
	The clear and simultaneous jumps in charge transfer and lattice dimerization is 
a manifestation of the strong coupling of charge and lattice, 
as expected in the regime of electronic ferroelectricity. 
	As the pressure increases, however, the jumps are diminished, 
indicating that charge and lattice become less coupled and eventually decoupled at $\sim$ 7.5 kbar 
(Fig.~\ref{Fig3}d). 
	The separation of the electronic and lattice instabilities is considered 
an inevitable process that makes the electronic ferroelectricity cross over into the ionic one.

	Electron transfer between different orbitals in adjacent molecules plays a primary role in 
the emergence of the electronic ferroelectricity of current interest, 
whereas the ionic displacement with electrons bound within respective molecules carries 
the conventional ferroelectricity. 
	The present experiments reveal that, in an organic ferroelectric material, TTF-QCl$_{4}$, 
the former transitions to the latter through the separation of the charge-transfer and 
lattice instabilities.
	The separation of the two instabilities possibly generates mobile neutral-ionic domain 
walls and solitons, which carry electric currents; indeed, recent theoretical 
studies \cite{Fukuyama_2016, Tsuchiizu_2016} suggested anomalous electric conductivity by such topological excitations.

	We thank H. Fukuyama, M. Ogata and N. Nagaosa for critical discussions. 
	This work was supported by the JSPS Grant-in-Aids for Scientific Research (S) 
(Grant No. 25220709) and for Scientific Research (C) (Grant No. 17K05846), 
and by CREST (Grant Number: JPMJCR1661), Japan Science and Technology Agency.

	R.T., K.S., F.I., and M.H. contributed equally to this work.


\begin{thebibliography}{99}
%Ref01
\bibitem{Resta_2007} R. Resta and D. Vanderbilt, in Phys. Ferroelectr. A Mod. Perspect., 
edited by K. M. Rabe, C. H. Ahn, and J.-M. Triscone (Springer, 2007), pp. 31-68.

%Ref02
\bibitem{Cointe_1995} M. Le Cointe, M. H. Lem$\acute{\textrm{e}}$e-Cailleau, H. Cailleau, 
B. Toudic, L. Toupet, G. Heger, F. Moussa, P. Schweiss, K. H. Kraft, and N. Karl, 
Phys. Rev. B \textbf{51}, 3374 (1995).

%Ref03
\bibitem{Kobayashi_2012} K. Kobayashi, S. Horiuchi, R. Kumai, F. Kagawa, Y. Murakami, 
and Y. Tokura, 
Phys. Rev. Lett. \textbf{108}, 237601 (2012).

%Ref04
\bibitem{Mayerle_1979} J. J. Mayerle, J. B. Torrance, and J. I. Crowley, 
Acta Crystallogr. Sect. B Struct. Crystallogr. Cryst. Chem. \textbf{35}, 2988 (1979).

%Ref05
\bibitem{Giovannetti_2009} G. Giovannetti, S. Kumar, A. Stroppa, J. Van Den Brink, and S. Picozzi, 
Phys. Rev. Lett. \textbf{103}, 266401 (2009).

%Ref06
\bibitem{Ishibashi_2014} S. Ishibashi and K. Terakura, J. Phys. Soc. Japan \textbf{83}, 73702 (2014).

%Ref07
\bibitem{Torrance_1981} J. B. Torrance, J. E. Vazquez, J. J. Mayerle, and V. Y. Lee, 
Phys. Rev. Lett. \textbf{46}, 253 (1981).

%Ref08
\bibitem{Torrance_1981b}J. B. Torrance, A. Girlando, J. J. Mayerle, J. I. Crowley, V. Y. Lee, 
P. Batail, and S. J. LaPlaca, Phys. Rev. Lett. \textbf{47}, 1747 (1981).

%Ref09
\bibitem{Girlando_1983} A. Girlando, F. Marzola, C. Pecile, and J. B. Torrance, 
J. Chem. Phys. \textbf{79}, 1075 (1983).

%Ref10
\bibitem{Tokura_1985} Y. Tokura, Y. Kaneko, H. Okamoto, S. Tanuma, T. Koda, T. Mitani, and G. Saito, 
Mol. Cryst. Liq. Cryst. \textbf{125}, 71 (1985).

%Ref11
\bibitem{King-Smith_1993} R. D. King-Smith and D. Vanderbilt, 
Phys. Rev. B \textbf{47}, 1651 (1993).

%Ref12
\bibitem{Vanderbilt_1993} D. Vanderbilt and R. D. King-Smith, 
Phys. Rev. B \textbf{48}, 4442 (1993).

%Ref13
\bibitem{Resta_1994} R. Resta, 
Rev. Mod. Phys. \textbf{66}, 899 (1994).

%Ref14
\bibitem{Takaoka_1987} K. Takaoka, Y. Kaneko, H. Okamoto, Y. Tokura, T. Koda, T. Mitani, 
and G. Saito, 
Phys. Rev. B \textbf{36}, 3884 (1987).

%Ref15
\bibitem{Cailleau_1997} M. Lem$\acute{\textrm{e}}$e-Cailleau, M. Le Cointe, H. Cailleau, 
T. Luty, F. Moussa, J. Roos, D. Brinkmann, B. Toudic, C. Ayache, and N. Karl, 
Phys. Rev. Lett. \textbf{79}, 1690 (1997).

%Ref16
\bibitem{Cointe_2017} M. Buron-Le Cointe, E. Collet, B. Toudic, P. Czarnecki, and H. Cailleau, 
Crystals \textbf{7}, 285 (2017).

%Ref17
\bibitem{Murata_1997} K. Murata, H. Yoshino, H. O. Yadav, Y. Honda, and N. Shirakawa, 
Rev. Sci. Instrum. \textbf{68}, 2490 (1997).

%Ref18
\bibitem{Gourdji_1991} M. Gourdji, L. Guib$\acute{\textrm{e}}$, A. P$\acute{\textrm{e}}$neau, 
J. Gallier, B. Toudic, and H. Cailleau, 
Solid State Commun. \textbf{77}, 609 (1991).

%Ref19
\bibitem{Gallier_1993} J. Gallier, B. Toudic, Y. Delugeard, H. Cailleau, M. Gourdji, A. Peneau, and 
L. Guibe, 
Phys. Rev. B \textbf{47}, 688 (1993).

%Ref20
\bibitem{Matsuzaki_2005} H. Matsuzaki, H. Takamatsu, H. Kishida, and H. Okamoto, 
J. Phys. Soc. Japan \textbf{74}, 2925 (2005).

%Ref21
\bibitem{Bayer_1951} H. Bayer, 
Zeitschrift F$\ddot{\textrm{u}}$r Phys. \textbf{130}, 227 (1951).

%Ref22
\bibitem{Koukoulas_1990} A. A. Koukoulas and M. A. Whitehead, 
Chem. Phys. Lett. \textbf{167}, 379 (1990).

%Ref23
\bibitem{Horiuchi_2000} S. Horiuchi, Y. Okimoto, R. Kumai, and Y. Tokura, 
J. Phys. Soc. Japan \textbf{69}, 1302 (2000).

%Ref24
\bibitem{Dengl_2014} A. Dengl, R. Beyer, T. Peterseim, T. Ivek, G. Untereiner, and M. Dressel, 
J. Chem. Phys. \textbf{140}, 244511 (2014).

%Ref25
\bibitem{Okamoto_1989} H. Okamoto, T. Koda, Y. Tokura, T. Mitani, and G. Saito, 
Phys. Rev. B \textbf{39}, 10693 (1989).

%Ref26
\bibitem{Fukuyama_2016} H. Fukuyama, and M. Ogata,
J. Phys. Soc. Japan \textbf{85}, 23702 (2016).

%Ref27
\bibitem{Tsuchiizu_2016} M. Tsuchiizu, H. Yoshioka,and H. Seo
J. Phys. Soc. Japan \textbf{85}, 104705 (2016).

\end{thebibliography}
\end{document}

\title{Supplementary Material for the Paper "Separation of Charge Instability and Lattice Symmetry Breaking in an Organic Ferroelectric"}

%\author{Ryosuke Takehara$^{1}$, Keishi Sunami$^{1}$, Fumitatsu Iwase$^{1}$, Masayuki Hosoda$^{1}$, 
%Kazuya Miyagawa$^{1}$, Tatsuya Miyamoto$^{2}$, Hiroshi Okamoto$^{2,3}$, and Kazushi Kanoda$^{1}$}
	
%\affiliation{
%$^{1}$ Department of Applied Physics, University of Tokyo, Bunkyo-ku, Tokyo, 113-8656, Japan \\
%$^{2}$ Department of Advanced Materials Science, University of Tokyo, Kashiwa, Chiba, 277-8561, Japa%n \\
%$^{3}$ AIST-UTokyo Advanced Operando-Measurement Technology 
%Open Innovation Laboratory (OPERANDO-OIL), 
%National Institute of Advanced Industrial Science and Technology (AIST), Chiba 277-8568, Japan}

%\date{\today}

\maketitle

\textbf{Line assignment of the NQR spectra.}
	In the neutral phase of TTF-QCl$_{4}$, the crystal structure is monoclinic $P2_{1}/n$ 
\cite{Cointe_1995}; the DADA arrangement is uniform (non-dimeric) and all the QCl$_{4}$ molecules 
are crystallographically equivalent. 
	Each molecule has two inequivalent Cl atoms (labelled A and B in Fig.~1a) 
because an inversion centre is located on the QCl$_{4}$ molecule despite the absence of 
intramolecular mirror-reflection symmetry in the crystal. 
	Thus, two NQR lines are observed in the neutral phase. 
	At ambient pressure, we observed two $^{35}$Cl NQR spectra at 37.54 and 36.96 MHz at 260 K, 
arising from the A and B sites. 
	Both spectra behaved similarly; namely, their positions shifted to higher frequencies 
with decreasing temperature and split at $T_{\textrm{c}} \sim$  82 K, 
reproducing the previous NQR results \cite{Gourdji_1991, Gallier_1993}. 
Thus, we investigated one of the two lines, residing on the higher frequency side.

\vspace{2ex}

\textbf{Estimation of the degree of charge transfer from the NQR frequency.} 
	The NQR frequency $\nu_{\textrm{Q}}$ is proportional to the electric-field gradient (EFG) 
at the nuclei and therefore is sensitive to the degree of charge transfer $\rho$. 
	Using the relationship between $\nu_{\textrm{Q}}$ and $\rho$, one can know the $\rho$ value from 
the $\nu_{\textrm{Q}}$ value. 
	However, there are two additional effects on $\nu_{\textrm{Q}}$: 
the temperature effect arising from the thermally activated molecular motions, 
which effectively reduces the anisotropy of the quadrupole tensor 
due to the motional averaging \cite{Bayer_1951, Koukoulas_1990}, 
and the pressure effect coming from the change of molecular packing, which can change the EFG. 
	Owing to these two effects, it is not straightforward to convert $\nu_{\textrm{Q}}$ to $\rho$. 
	Here, we describe our method for estimating $\rho$ from $\nu_{\textrm{Q}}$ in TTF-QCl$_{4}$, 
considering these effects.

\begin{figure}
	\includegraphics[width=8.6cm]{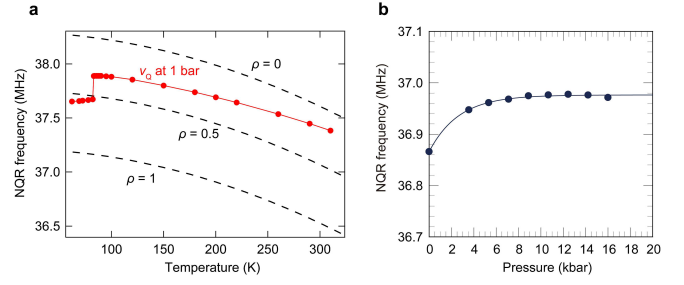}
	\caption{\label{FigS1} (Color online) 
Relationship between $^{35}$Cl NQR frequency and the degree of charge transfer $\rho$. 
	a, Temperature dependence of $^{35}$Cl NQR frequency. 
	The red points indicate the experimental $\nu_{\textrm{Q}}$ data at 1 bar; 
the red line is a guide for the eye. 
	The broken lines represent the calculated temperature dependences of 
NQR frequency for $\rho$ = 0 (upper), 0.5 (middle) and 1 (lower). 
	b, Pressure dependence of $^{35}$Cl NQR frequency in neutral QCl$_{4}$ crystals. 
	The line is a fit of the form $a[1-\exp(-bP)]$ to the experimental data of 
$\Delta\nu_{\textrm{Q}}^{\textrm{QCl4}}(P)=\nu_{\textrm{Q}}^{\textrm{QCl4}}(P)-\nu_{\textrm{Q}}^{\textrm{QCl4}}(0)$, which yields $a$ = 0.110 MHz and $b$ = 0.379 kbar$^{-1}$.}
\end{figure}

	First, we describe our analysis of the $\nu_{\textrm{Q}}$ data at ambient pressure. 
	We make three assumptions: 
(i) at a given temperature, $\nu_{\textrm{Q}}$ depends linearly on $\rho$; 
(ii) the proportionality coefficient $A$ between the variations of $\nu_{\textrm{Q}}$ and 
$\rho$ is independent of temperature; and 
(iii) the temperature dependence of $\nu_{\textrm{Q}}$ in 
the state of $\rho=0$, $\nu_{\textrm{Q}}^{\rho =0}$, obeys the empirical formula given by 
Koukoulas \textit{et al.},$\nu_{\textrm{Q}}^{\rho =0}(T)=\nu_{0}$exp$(-\alpha T^{2})$, 
where $\nu_{0}$ and $\alpha$ are constants \cite{Koukoulas_1990}. 
	These assumptions are formulated as $\nu_{\textrm{Q}}(T)=A\rho(T)+\nu_{0}$exp$(-\alpha T^{2})$. 
	The value of $A$ is estimated at -1.08 MHz/$\rho$ from the jump in NQR frequency 
at the ferroelectric transition, $\Delta \nu_{\textrm{Q}}(T_{\textrm{c}})$ = 0.216 MHz, 
and the change of $\rho$ evaluated from the infrared spectroscopy measurements, 
$\Delta \rho^{\textrm{IR}}(T_{\textrm{c}})$ = 0.20 \cite{Horiuchi_2000}. 
	Then, we use the $\nu_{\textrm{Q}}$ and $\rho^{\textrm{IR}}$ values at two temperatures, 
$T$ = 296 K ( $\nu_{\textrm{Q}}$ = 37.624 MHz and $\rho^{\textrm{IR}}$ = 0.18) and 
$T_{\textrm{c}}^{+}$ (the lowest temperature in the high-temperature phase, 
$\nu_{\textrm{Q}}$ = 37.888 MHz and $\rho^{\textrm{IR}}$ = 0.33) 
\cite{Matsuzaki_2005, Horiuchi_2000} to determine the $\nu_{0}$ and $\alpha$ values, 
which yields $\nu_{0}$ = 38.296 MHz and $\alpha$ = 2.02 $\times$ 10$^{-7}$ K$^{-2}$. 
	Thus, the deduced relationship between $\nu_{\textrm{Q}}$ and $\rho$ at ambient pressure 
is depicted in Fig.~S\ref{FigS1}a.

	Next, to extract the $\rho$ values under various pressures, 
we further assume that (iv) $A$, $\nu_{0}$, and $\alpha$ are independent of pressure and 
(v) the pressure effect is incorporated by adding a correction term, $\Delta\nu_{\textrm{Q}}(P)$, 
to the $\nu_{\textrm{Q}}(T)$ for any temperature. 
	For $\Delta\nu_{\textrm{Q}}(P)$, we refer to the pressure dependence of the $^{35}$Cl NQR 
frequency of the neutral QCl$_{4}$ crystal, $\nu_{\textrm{Q}}^{\textrm{QCl4}}$, 
in which the degree of charge transfer is invariant under pressure variation. 
	Figure~S\ref{FigS1}b shows the pressure dependence of $\nu_{\textrm{Q}}^{\textrm{QCl4}}$ 
at room temperature, which slightly increases with pressure up to 8 kbar and saturates 
at higher pressures. 
	The pressure dependence, 
$\Delta\nu_{\textrm{Q}}^{\textrm{QCl4}}(P)= \nu_{\textrm{Q}}^{\textrm{QCl4}}(P)-\nu_{\textrm{Q}}^{\textrm{QCl4}}(0)$ , is fitted by the equation 
$\Delta\nu_{\textrm{Q}}^{\textrm{QCl4}}=a[1-\exp(-bP)]$ with $a$ = 0.110 MHz and 
$b$ = 0.379 kbar$^{-1}$. 
	Using this equation for the pressure effect, the relationship between $\nu_{\textrm{Q}}$ and 
$\rho$ is eventually expressed as $\nu_{\textrm{Q}}=A\rho(T)+\nu_{0}\exp(-\alpha T^{2})
+a[1-\textrm{exp}(-bP)]$. 
	Using this formula, we obtained the temperature dependence of $\rho$ under 
various pressures as shown in Fig. 3c in the main text. The plots of $\rho$ against temperature 
and pressure are qualitatively consistent with those of earlier optical measurements \cite{Dengl_2014}.

%\newpage